\documentclass[a4paper]{article}

\usepackage{INTERSPEECH2018}

\title{An Ensemble SVM-based Approach for Voice Activity Detection}
\name{Jayanta~Dey$^1$, Md.~Sanzid~Bin~Hossain$^2$, Mohammad~Ariful~Haque$^3*$}
\address{
Dept. of Electrical and Electronic Engineering, Bangladesh University of Engineering and Technology}
\email{deyjayanta76@gmail.com, sanzidbinhossain33@gmail.com, arifulhoque@eee.buet.ac.bd}

\begin{document}

\maketitle
\begin{abstract}
Voice activity detection (VAD), used as the front end of speech enhancement, speech and speaker recognition algorithms, determines the overall accuracy and efficiency of the algorithms. Therefore, a VAD with low complexity and high accuracy is highly desirable for speech processing applications. In this paper, we propose a novel training method on large dataset for supervised learning-based VAD system using support vector machine (SVM). Despite of high classification accuracy of support vector machines (SVM), trivial SVM is not suitable for classification of large data sets needed for a good VAD system because of high training complexity. To overcome this problem, a novel ensemble-based approach using SVM has been proposed in this paper. The performance of the proposed ensemble structure has been compared with a The performance of the proposed ensemble structure has been compared with a feedforward neural network (NN). Although NN performs better than single SVM-based VAD trained on a small portion of the training data, ensemble SVM gives accuracy comparable to neural network-based VAD. Ensemble SVM and NN give $88.74$\% and $86.28$\% accuracy respectively whereas the stand-alone SVM shows $57.05$\% accuracy on average on the test dataset.
\end{abstract}


\section{Introduction}
Voice activity detection is basically the act of separating the speech and the non-speech portions of an audio recording. As a typical speech signal may contain silence, mono-tonic noise and  even music frames, sorting out the speech frames from the recording using an efficient VAD in the front end is of great importance for reliable operation of the speech processing algorithms.

A number of algorithms for voice activity detection have been proposed in the literature. Among them frame energy, periodicity measure \cite{tucker1992voice} or entropy-based \cite{renevey2001entropy} methods are elegant in the sense of their simplicity and time-efficiency. However, they are based on parameters selection that are tuned for a particular situation and can not separate more critical non-speech frames like music accurately. A solution to this problem can be found by adopting relatively complex statistical approach such as statistical hypothesis testing \cite{chang2006voice}, long-term spectral divergence measure \cite{ramirez2004efficient}, amplitude probability distribution \cite{tanyer2000voice} and low-variance spectrum estimation \cite{davis2006statistical}. However, these methods need to estimate the background noise level and are also prone to several parameters tuning. More recent studies attempts to solve the problem of VAD from machine learning point-of-view \cite{shin2010voice}, \cite{wu2011maximum}, \cite{zhang2013deep} that classifies an audio frame as speech or non-speech. The main problem associated with these approaches is that they need to be trained on a large dataset which includes a rich non-speech instances for a satisfactory efficiency. This poses a problem for learners such as SVM that has high classification accuracy and yet can not be trained on a large dataset due to complex training algorithm.

In this paper, we propose a novel ensemble technique to train SVM learners on a large dataset for voice activity detection and compare it with the performance of a neural network-based classifier trained on the similar dataset. Time efficiency is a crucial factor for VAD implementation and unlike the other methods reported in the literature \cite{van2013robust} that use a large set of features, we focus our efforts on building classifier based on MFCC features only. These MFCC features are popularly used for speech processing algorithms. Therefore, the extracted features can be used in the subsequent stages making the overall procedure more efficient. In our work, we have trained a number of SVMs on non-overlapping small datasets to cover the whole large dataset and their predicted probability is used as features for the output layer SVM that gives the final decision. The proposed SVM-ensemble gives approximately similar accuracy compared to the state of the art neural network \cite{van2013robust}. This approach shows significant improvement in terms of accuracy from the stand-alone SVM and also the variance in result for a single SVM is smoothed out by the ensemble.

The paper is organized as follows. Section II describes problem formulation and data description. The system architecture is discussed in section III and the system efficacy is established in section IV. Finally, mentioning the contributions and our future work, the paper is concluded in section VI.
 
\begin{figure*}[ht!]
\centering
\includegraphics[width = 6 in, height = 3.4 in]{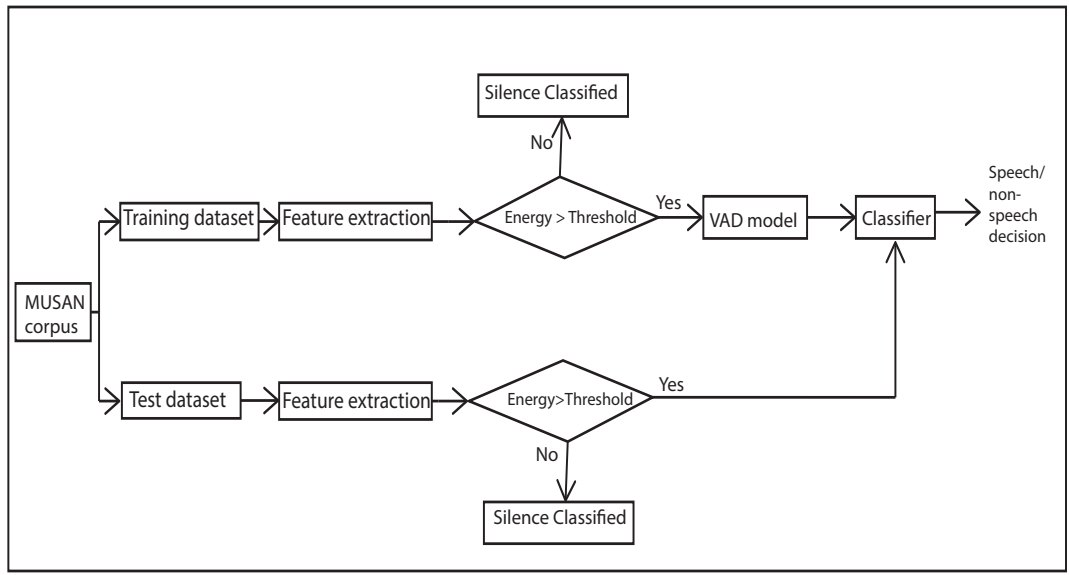}
\caption{General system model for the proposed voice activity detection.}
\label{vad_model}
\end{figure*}

\section{Problem Formulation and Data Description}
In our work, we have used \textbf{MUSAN} corpus \cite{snyder2015musan} as both our testing and training database. The corpus consists of approximately $109$ hours of speech, music and noise data that makes it an ideal database to be used for supervised learner-based VAD application. The speech dataset contains silent frames which are excluded from the training dataset by a log-mel energy based thresholding described later in the paper and a test dataset of duration of about $3$ hours has been separated from the corpus where the silent frames of speech recordings were annotated by a human listener. Now both the training and testing dataset were divided into frames of duration $25$ ms with $15$ ms overlap. Therefore, the challenge is to build a learner trained on the features extracted from the training frames that can classify the testing frames as speech or non-speech. Here we have used an ensemble-based approach for training an SVM learner and compared it with a trivial architecture-based neural network.

\begin{figure*}[ht!]
\centering
\includegraphics[width = 6.5 in, height = 3.3 in]{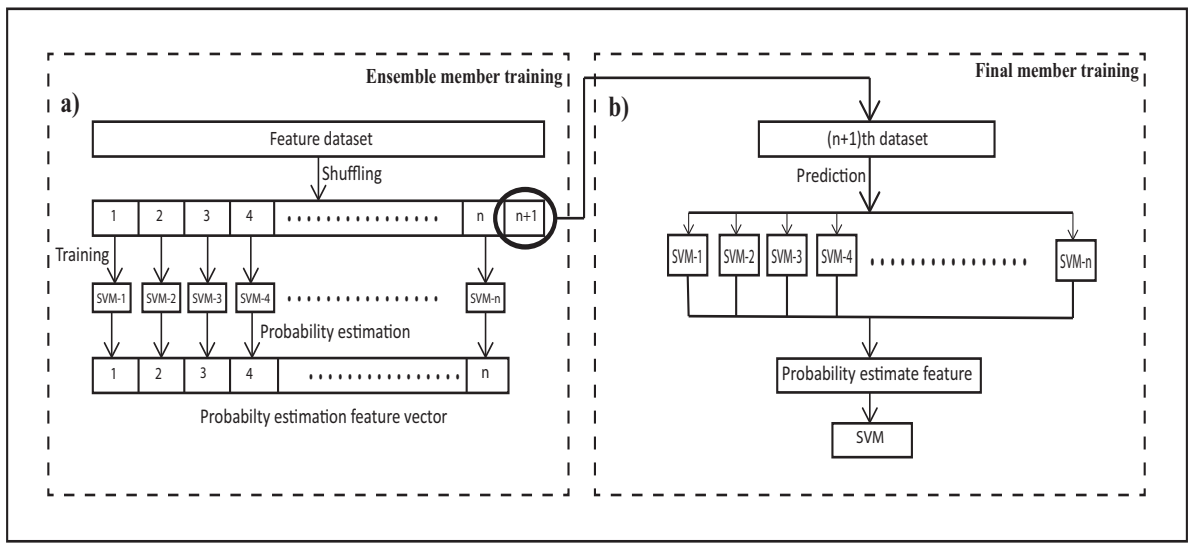}
 \caption{a) Training $n$ numbers of SVM for ensembling, b) Training the final output layer SVM.}
 \label{ensemble_model}
\end{figure*}

\begin{figure*}[ht!]
\centering
\includegraphics[width = 6.5 in, height = 2.8 in]{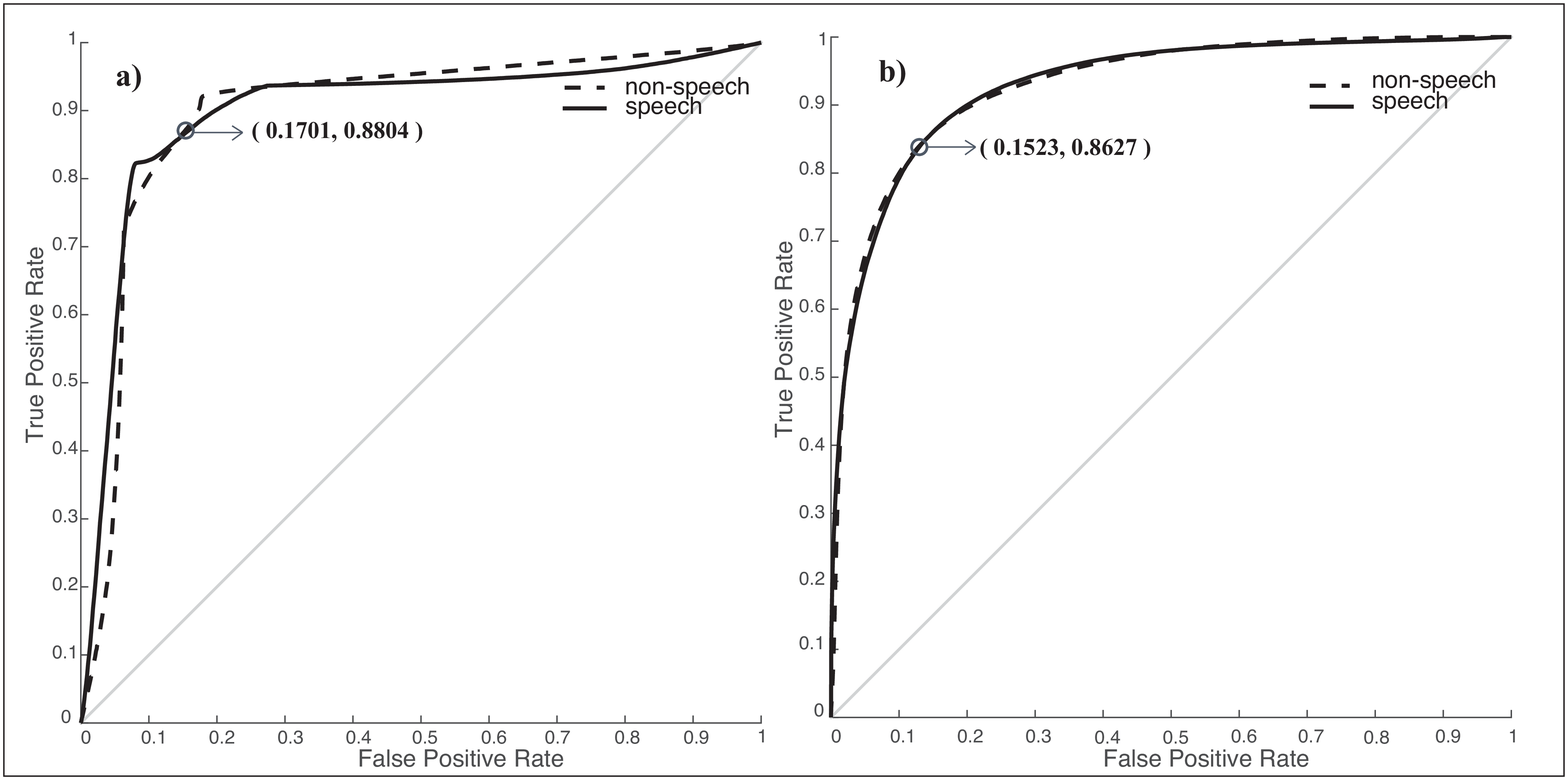}
 \caption{a) ROC curves for ensemble-SVM, b) ROC curves for NN.}
 \label{ROC}
\end{figure*}

\section{VAD System Description}
In this work, we aim to develop a binary classification system in which one class consists of only speech and the other one contains silence, music and noise. A general overview of the VAD system is shown in Fig. \ref{vad_model}. As the silent frames can be separated quite efficiently by energy-based thresholding only, they were excluded from the training data and detected in the testing phase using a thresholding on the log-mel energy that is the first MFCC feature. Then the remaining frames were classified using a supervised-learner-based VAD system.

\subsection{Feature Extraction}
In order to train the classifier, we have used MFCC features as they are the standard features used in speech processing. There are many software packages to extract the MFCC features efficiently and hence it will be an elegant feature-set to be used in VAD where time-efficiency is highly desirable. MFCC is a spectral feature inspired by human auditory model. As human ear is efficient in distinguishing between speech and non-speech, we hope the MFCC features will also be effective for our purpose. Here we have used $13$ MFCC features for a particular audio frame.

\subsection{Classification using Neural Network}
We have developed and evaluated our deep neural network (NN) model with python library Keras$\textsuperscript{\textregistered}$ and numerical computation software library  tensorflow. We have trained a fully connected  neural network model with an input layer of $13$ input variable ,two hidden layers with $12$ and $8$ neurons respectively and an output layer with one neuron. We have initialized network weights to small random numbers which was generated from a uniform distribution. Details of the layer architecture is given in Table \ref{tab: table1}. We have used binary crossentropy as loss function and gradient decent algorithm ‘Adam’ as optimizer. 
\begin{table}[ht!]
\caption{Architecture and Training Parameters of the Neural Network}
\label{tab: table1}
\resizebox{\columnwidth}{!}{
\begin{tabular}{||c||c|c|c|c|c|}\hline
\multicolumn{1}{||c||}{ } &
\multicolumn{1}{c|}{ } &
\multicolumn{1}{c|}{ } & 
\multicolumn{1}{c|}{ } & 
\multicolumn{1}{c|}{ } &
\multicolumn{1}{c|}{ } \\ 
 Layer Details & Node Number & Activation & Epochs & Batch Size & Validation Split\\ \hline
Hidden Layer 1 & 12 & relu & & & \\ \cline{1-3}
Hidden Layer 2 & 8 & relu & 100 & 100 & .2\\ \cline{1-3}
Output Layer & 1  & sigmoid & & & \\ \hline
\end{tabular}
}
\end{table}

\subsection{Classification using SVM}
In our work, we have used libsvm toolbox \cite{chang2011libsvm} for SVM-based classification. As a stand-alone SVM is not suitable for the large train dataset described earlier, we shuffle the feature sets obtained from the train dataset and divide it into $n$ non-overlapping smaller datasets. The overview of the procedure is shown in the Fig. \ref{ensemble_model}. Here we have used an ensemble of $n = 5$ SVMs with non-linear rbf kernel. The values of gamma and C parameters of the kernel have been tuned from the $2$-fold cross-validation accuracy using grid-search \cite{hsu2003practical}. The rationale behind choosing an ensemble of $5$ learners is that the performance saturates for higher number of learners. In the first stage, the feature dataset was shuffled and divided into $6$ portions. Among them the first $5$ segments has been used for training the ensemble-members and then each of the trained members gave probability estimates for the held-out $6$-th portion of the dataset. For one feature vector of $13$ MFCC features each of the members gives $1$ probability estimate and thus a new feature space of $5$ features have been derived from one input feature vector which makes the procedure completely data-driven without using any heuristic thresholding. These feature vectors have been used for training an SVM classifier with rbf kernel in the final layer. Here instead of using a majority voting based decision, we have used SVM as majority voting system may give estimate biased to a particular ensemble member without considering the other members. In case of larger dataset even than that of used in this work, the number of SVM layers in the system may be increased similar to an NN architecture.

\section{Result Analysis}
The proposed ensemble-SVM was tested on the test dataset described in the data description section. To prove the efficacy of the classifier, we attempt to examine the ensemble architecture from different perspectives such as effect of layer members, stability and finally we compare the classifier with a stable, efficient neural network. In order to test the effect of ensembling, we gradually increase the number of ensemble member and observe their accuracy in Table \ref{tab: table 2}.

\begin{table}[ht!]
\caption{Effect on performance for increasing ensemble members}
\label{tab: table 2}
\resizebox{\columnwidth}{!}{
\begin{tabular}{||c||c|c|}\hline
\multicolumn{1}{||c||}{ } &
\multicolumn{1}{c|}{ } & 
\multicolumn{1}{c|}{ } \\
 Ensemble Member No. & Accuracy & Average individual member accuracy\\ \hline
1 & 57.05\% & \\ \cline{1-2}
2 & 72.25\% & \\ \cline{1-2}
3 & 78.32\% & 57.05\%\\ \cline{1-2}
4 & 87.02\% & \\ \cline{1-2}
5 & 88.74\% & \\ \cline{1-2}
6 & 88.82\% & \\ \hline
\end{tabular}
}
\end{table}

From Table \ref{tab: table 2} we see that the accuracy of classification increases for increasing number of ensemble members and the accuracy almost saturates after $n = 5$ ensemble members. Although individual SVM trained on smaller dataset shows poor performance of $57.05\%$, the effect of ensembling is evident from the higher classification accuracies of the ensembles. Again to observe estimation variance reduction of the ensemble, we present the accuracy of two testing files in Table \ref{tab: table 3}. For example, in the case of `speech-librivox-0011' file, if we use stand-alone SVM the accuracy may vary in the range $8.65\sim 91.35\%$ as the stand-alone SVMs are trained on different portions of data and hence, they may perform differently for a particular test-case. From Table \ref{tab: table 3} we can observe that the accuracy of individual SVM may fluctuate whereas their ensemble accuracy remains stable and close to the maximum individual accuracy.

\begin{table}[ht!]
\caption{Effect on performance for increasing ensemble members}
\label{tab: table 3}
\resizebox{\columnwidth}{!}{
\begin{tabular}{||c||c|c|}\hline
\multicolumn{1}{||c||}{ } &
\multicolumn{1}{c|}{ } & 
\multicolumn{1}{c|}{ } \\
File Name & Ensemble Member Accuracy & Total Ensemble Accuracy\\ \hline
 & 8.6466\% & \\ \cline{2-2}
 & 8.6466\% & \\ \cline{2-2}
speech-librivox-0011 & 91.3534\% & 90.2256\%\\ \cline{2-2}
 & 39.4737\% & \\ \cline{2-2}
 & 18.797\% & \\ \hline
 & 5.26316\% & \\ \cline{2-2}
 & 30.8271\% & \\ \cline{2-2}
noise-sound-bible-0031 & 21.4286\% & 72.45\%\\ \cline{2-2}
    & 49.6241\% & \\ \cline{2-2}
     & 68.797\% & \\ \hline
\end{tabular}
}
\end{table}

Finally, we compare the performance of the ensemble SVM with the NN described in the subsection $3.2$. NN and ensemble-SVM give $86.28$\% and $88.74$\% accuracy respectively. Their ROC curves are given in Fig. \ref{ROC}. The operating point of each classifier is shown by using a circle on the curves which shows that the ensemble-SVM has a better true positive rate of $88.04\%$ and a slightly high false positive rate of $17.01\%$ compared to that of NN. The average area under curve (AUC) for NN and ensemble-SVM are $0.9284$ and $0.9167$ respectively. From these performance indices, we can conclude that their performances are comparable.

\section{Conclusion}
In this paper, we have proposed a novel ensemble SVM-based approach for voice activity detection. The efficacy of the proposed ensembling method has been established in the result section through comparing with NN and testing on the test dataset. Here the member SVMs are independent of each other and hence they can operate parallelly resulting in a significant reduction in runtime. Again different ensemble member can be trained on different types of feature giving a more robust VAD system. Replacing some of the layers of an NN with SVM layer may result in improved accuracy. In our future work, we will explore composite structure of NN and SVM with reduced training complexity.

\bibliographystyle{IEEEtran}

\bibliography{jd}

\end{document}